\documentclass[10pt,letterpaper,conference]{IEEEtran}
\IEEEoverridecommandlockouts
\ifCLASSINFOpdf
  \usepackage[pdftex]{graphicx}
\else
  \usepackage[dvips]{graphicx}
\fi

\usepackage[cmex10]{amsmath}
\usepackage{amssymb}
\usepackage{soul}
\usepackage{flushend,cite}
\usepackage{multirow}
\usepackage{subfigure}
\usepackage{multirow}

\def\draftmode{}
\ifx\draftmode\undefined
\newcommand{\comment}[1]{}
\else
\newcommand{\comment}[1]{ \marginpar{$\Longleftarrow$}{\bf $<$#1$>$} }
\fi

\begin{document}

\title{Practical Provably Secure Multi-node Communication}

\author{\IEEEauthorblockN{Omar Ali}
\IEEEauthorblockA{Wireless Research Center\\
E-JUST, Egypt\\
omar.ali@ejust.edu.eg} \and
\IEEEauthorblockN{Mahmoud F. Ayoub}
\IEEEauthorblockA{Dept. of Comp. and Sys. Eng.\\
Alexandria Univ., Egypt\\
mfayoub@alexu.edu.eg} \and \IEEEauthorblockN{Moustafa
Youssef}
\IEEEauthorblockA{Wireless Research Center\\
E-JUST, Egypt\\
moustafa.youssef@ejust.edu.eg}
}

\maketitle
\begin{abstract}
We present a practical and provably-secure multi-node communication scheme in
the presence of a passive eavesdropper. The scheme is based on a random
scheduling approach that hides the identity of the transmitter from the
eavesdropper. This random scheduling leads to ambiguity at the eavesdropper with
regard to the origin of the transmitted frame. We present the details of the
technique and analyze it to quantify the secrecy-fairness-overhead trade-off.
Implementation of the scheme over Crossbow Telosb motes, equipped with CC2420
radio chips, shows that the scheme can achieve significant secrecy gain with
vanishing outage probability. In addition, it has significant overhead advantage over direct extensions to two-nodes schemes. The
technique also has the advantage of allowing inactive nodes to leverage sleep
mode to further save energy.
\end{abstract}

\section{Introduction}
With the continuous growth of wireless networks and emerging new technologies
such as WiMAX and LTE, wireless networks security has received extensive
attention. Current popular security schemes, e.g. public key cryptography, are
based on computationally secure trapdoor one-way functions \cite{Faqs02}. These
schemes depend on the assumption that it is \textit{hard} for an attacker to
decipher the message without knowing the trapdoor (i.e. the secret key).
However, these schemes do not prevent a computationally unlimited attacker from
decrypting the message without knowing the trapdoor as it is not proven yet that
one-way functions cannot be inverted efficiently \cite{Faqs02}. Therefore, these
schemes are not \textit{provably secure}.

Information theoretic secrecy, on the other hand, introduces the possibility of
having perfectly secure communication independently from the computational
capabilities of the attacker \cite{Shannon49,Wyner75,Maurer93,Gopala06,Tang07,Omar09, Sabagh10, Sabagh11,forensic11,Elmorsy,Arora2009,karim_j,aly_j,multinode_demo}. In particular, Shannon \cite{Shannon49} proved
that, using a shared secret key $K$,  the achievability of \emph{perfect
secrecy} requires that the entropy of $K$ be at least equal to the entropy of
the message $M$ (i.e., $H(K) \geq H(M)$). Wyner showed that it is possible to
send perfectly secure messages at a non-zero rate, \textit{without} relying on
secret keys or any limiting assumptions on the computational power of the
wiretapper, under the condition that the source-wiretapper channel is a degraded
version of the source-destination channel \cite{Wyner75}. This was later
extended to the non-degraded scenario in \cite{Maurer93}. In
\cite{Gopala06,Tang07}, the effect of fading on the secrecy capacity was studied
and it was shown that distributing the message across different fading
realizations actually increases the secrecy capacity.

Although information theoretic security schemes provide provable security, they
have been considered \textbf{not practical} due to the simplifying assumptions
they have to prove their security. Recently, we have introduced a number
of practical and provably-secure protocols for \textbf{\emph{two-node}}
communication based on information theoretic concepts. Our work in \cite{Omar09,
Sabagh10, Sabagh11,forensic11} exploits the multi-path nature of the wireless
medium to provide \textbf{practical} information-theoretic security in channels
with feedback. The basic idea is to distribute the secret key among multiple ARQ
frames. This concept has been used to enhance the security of practical Wi-Fi
and RFID protocols at the expense of slight loss in throughput.

Direct extensions of these two-node schemes to the \textbf{multi-node} case, by applying
the protocol to each pair of communicating nodes, lead to a considerable waste
of throughput. This is due to optimizing each pair independently, extending the
two-node overhead to the multi-node case.

In this paper, we present a practical and provably-secure scheme at the presence
of a passive eavesdropper that is designed for the multi-node case from the
beginning. Our scheme is based on a novel two-phase approach: in the first
phase, i.e. the selection phase, a node is selected as the transmitter using
information theoretic techniques that hide the identity of the selected node. In
the second phase, i.e. the data transmission phase, data frames are transmitted
without the source/destination ID in the packet header. This leads to ambiguity at the
eavesdropper. The length of the data transmission phase can be tuned to trade-off
secrecy and efficiency. Nodes not selected at the selection phase can sleep to
the next cycle, further reducing their energy consumption. We present different
variations of the basic scheme, all having the same overhead, that can achieve
different secrecy-fairness trade-offs.  We evaluate our proposed schemes both
analytically and through implementation over Crossbow Telosb motes, equipped
with CC2420 radio chips. Our evaluation shows that the scheme can achieve
both significant secrecy gain and decrease in overhead as compared to direct extensions to the two-node schemes.

The rest of the paper is organized as follows: We define the system model in
Section~\ref{sec:model}. Section \ref{sec:basic} presents the basic scheme. In
Section~\ref{sec:extended}, we present four different extensions to the basic
scheme that can achieve different secrecy-fairness trade-offs. We analyze the
proposed schemes in Section~\ref{sec:analysis} along with the system implementation. We finally
conclude the paper in
Section~\ref{sec:conclude}.

\section{System Model}
\label{sec:model}

\begin{figure}
\centering
\includegraphics[width=0.5\linewidth]{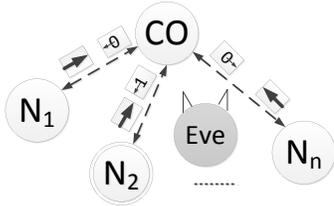}
\caption{Network model. A star topology is assumed where each node have to
send/receive through the coordinator (CO). In the selection phase, a node is selected if it receives a 1 bit from the CO using dialog codes($N_2$ in this
example).}
\label{fig:Network}
\end{figure}

We consider a network with $n$ legitimate nodes in the presence of a passive
eavesdropper (Eve). We assume a star topology, where all the traffic between nodes has
to go through a central node, i.e. a coordinator. This is common in WLANs,
cellular, and sensor networks\footnote{Note that a coordinator can also be
selected in a distributed manner if no central node is available.}. This
coordinator (e.g. access point, base station, or gateway) is responsible for
controlling the transmission in the network and assigning turns. All nodes are
equipped with half-duplex antennas. We further assume a time-slotted
communication system, where all nodes are synchronized
(Figure~\ref{fig:Network}).

For space constraints, we also assume that all nodes have equal load and Eve
cannot differentiate between nodes based on power\footnote{Power randomization
can be used in this case to confuse Eve as in our previous work
for two nodes \cite{Elmorsy}.}. We leave the general case to a future paper.

To further remove the need of acknowledgment, each message $M$ is erasure-coded
into $m$ frames such that the reception of any $k < m$ frames at the receiver
can be used to reconstitute $M$ with high probability. Note that using erasure
coding does not give any advantage to Eve as (a) she cannot determine the
identity of the transmitter and (b) there is no message level error detection
(only CRC at the frame level).

All system parameters are assumed to be known to the eavesdropper along with the
details of the technique, but not the instantaneous random values.

Each node needs to send and receive $k$ frames, each of $l$ bits. 
Table~\ref{Symbol_table} summarizes the
different symbols we use in the paper.

\begin{table}[!t]
\begin{tabular}{|c|l|l|}
\hline
Par. & Meaning & Default val. \\
\hline
\hline
$n$  & Number of network nodes. & 8\tabularnewline
\hline
$f$  & Num. of packets transm. in a single session. & 4\tabularnewline
\hline
$l$  & Frame length. & 1024\tabularnewline
\hline
$k$  & Num. of frames needed to reconstruct orig.
message. & 32\tabularnewline
\hline
$t$  & preamble length & 6\tabularnewline
\hline
\end{tabular}
\caption{Symbols used in the paper.}
\label{Symbol_table}
\end{table}

\section{Basic Scheme}
\label{sec:basic}

The basic idea of our approach is to hide the identity of the packet receiver at
any point in time from the eavesdropper. This creates an ambiguity that gives
advantage to legitimate nodes. In particular, the system works in two phases: a
receiver selection phase and a data transmission phase. In the first phase, the
coordinator (CO) informs one of the $n$ nodes securely, using physical layer security
principles, that it is selected as the receiver for the packets in the second
phase. In the data transmission phase, only the intended receiver has to be
active, all other nodes can go to sleep mode, reducing the energy consumption of
other nodes. An eavesdropper receiving a packet cannot determine the destination
of the packet, and hence has to guess.

The selection phase can be repeated as frequently as needed, by reducing the
number of frames during the data transmission phase ($f$), to increase security.
However, this increases the system overhead. Therefore, we have a tradeoff
between secrecy and overhead.

In the rest of this section, we present the details of the two phases, assuming
one way communication from the coordinator to the nodes. We present the
two-way case in Section~\ref{sec:extended} and analyze the overhead and security of
the system in Section~\ref{sec:analysis}.

\subsection{Receiver Selection Phase}

During this phase, the coordinator (CO) selects one of the $n$ nodes to be active during the data transmission phase. This is achieved by sending a one bit from the coordinator to each node indicating whether this node is selected (bit= 1) or not (bit= 0). We adopt dialog codes \cite{Arora2009} as the provably secure technique for exchanging these bits. In perfect conditions, dialog codes encode each bit as two bits and the receiver jams only one of these two bits randomly. Under a binary modulo-2 additive channel model, Eve has to guess which bit has been jammed, while Bob knows the correct bit with no ambiguity.

To accommodate realistic situations, \cite{Arora2009} adds a $t-1$ bit randomly
chosen preamble to the source bit and then encodes the result by dialog codes. The probability of Eve correctly guessing the transmitted bit ($P_E$) then becomes:

\begin{equation}
P_E= \frac{1}{2}\left ( 1 + \left ( 1-w \right )^{\frac{t+1}{2}} \right ), w = min (p, q)
\label{eq:preamble}
\end{equation}

where $p$ and $q$ are the probability of corrupting 0 to 1 and 1 to 0 respectively.

Equation~\ref{eq:preamble} shows that $P_E$ converges to $\frac{1}{2}$ as $t$ increases. \cite{Arora2009} showed that, in practice, the convergence speed is
much faster than $\frac{1}{2}\left ( 1 + \left ( 1-w \right )^{\frac{t+1}{2}} \right )$. In addition, they showed that for a typical environment, a \textbf{6 to 8 bit preamble is enough to confuse Eve}. This has been confirmed in our experiment.

\subsection{Data Transmission Phase}

During this phase, the selected node can send/receive $f$ data frames, where $f$
is a parameter that can be used to tune security versus overhead. Only the
selected node needs to be active. Transmitted frames do not contain the ID of
the receiver, which leads to ambiguity at the eavesdropper about the origin of
the frames. 

\begin{table*}[!t]
\centering
\caption{Comparison between the different scheduling schemes.}
\begin{tabular}{|p{3cm}||l|p{2cm}|p{2cm}|p{2cm}|p{2cm}|l|}
\hline
Scheme  & Abbrev. & Node selection fairness & From/To coordinator (direction)
fairness& One node outage prob. ($Pr_{\textrm{Out}}(n)$)& Network outage prob.
($Pr_{\textrm{NOut}}(n)$)& Overhead\\
\hline
\hline
Random node selection and random direction division & RN-RD & Long term & Long
term &
$\left.\frac{1}{\binom{\frac{2nk}{f}}{\frac{2k}{f}}}\right.{\frac{1}{\binom{2k}{
k}}}$ & \multirow{4}{*}{$\prod_{i=n}^{1}\Pr_{\textrm{out}}(i)$}  &
\multirow{4}{*}{$\frac{t(n+f)}{fl+t(n+f)}$}\\
\cline{1-5}
Random node selection and fair direction division & RN-FD & Long term & Short
term &
$\left.\frac{1}{\binom{\frac{2nk}{f}}{\frac{2k}{f}}}\right.{\frac{1}{\binom{f}{
\frac{f}{2}}^{\frac{2k}{f}}}}$ & & \\
\cline{1-5}
Fair node selection and random direction division & FN-RD &
Short term & Long term &
$\left.{\left(\frac{1}{n}\right)^{\frac{2k}{f}}}\right.{\frac{1}{\binom{2k}{k}}}
$ &  & \\
\cline{1-5}
Random within round node selection and fair direction division & FN-FD & Short
term & Short term &
$\left.\frac{1}{\binom{\frac{2nk}{f}}{\frac{2k}{f}}}\right.{\frac{1}{\binom{2k}{
k}}}$ &  & \\
\hline
\end{tabular}
\label{tab:schemes}
\end{table*}

\section{Two-Way Communication}
\label{sec:extended}
To allow for two-way communication, we need to specify which slots within a
session in the data transmission phase will be from/to the coordinator. In order
to do that, we add a third short direction determination phase between the
selection and data transmission phases in which the CO sends $f$ bits, using dialog codes again,
where each bit corresponds to a slot in the data transmission session. A bit set to 1 (0) corresponds to a from (to) CO slot. Note that the node ID is not sent in this phase. Therefore, Eve cannot
know the identity of the selected node.

In the data transmission phase, a node will follow the schedule received during
the direction selection phase.

In the rest of this section, we present four different schemes for assigning the
schedule between the different nodes and from/to the coordinator. The different
schemes can achieve different fairness-security goals as we quantify in
Section~\ref{sec:analysis}. Fairness refers to balancing the access opportunity
within nodes and between the from/to coordinator traffic. Therefore, we have
four combinations of fairness: node fairness (short and long term) and direction
fairness (short and long term). In all schemes, all nodes have to finish one
message of transmission before any node can start a new message for fairness purposes.
Table~\ref{tab:schemes} compares the different schemes.

We start by some notations followed by the details of the four schemes.

\subsection{Notations}
\begin{figure*}[!t]
\centering
\subfigure[Fair node selection  and fair direction division (FN-FD).]{
\includegraphics[height=0.16\textwidth,width=0.25\linewidth]{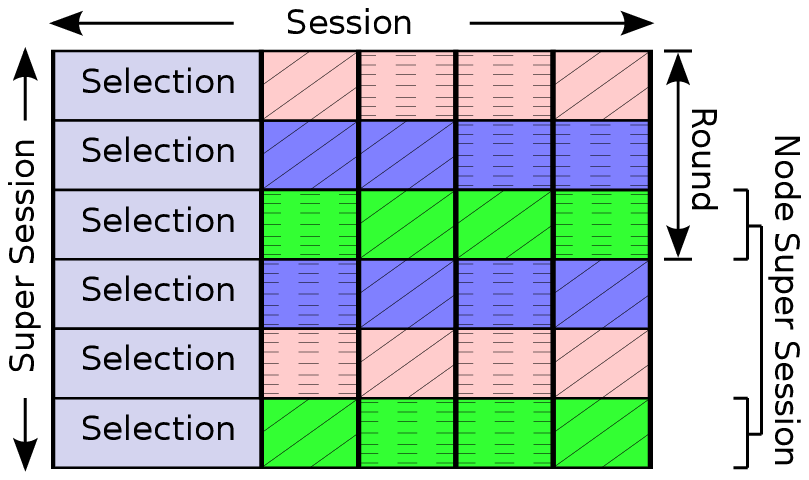}
\label{fig:FN_fd}
}\subfigure[Random node selection and fair direction division (RN-FD).]{
\includegraphics[height=0.16\textwidth,width=0.23\linewidth]{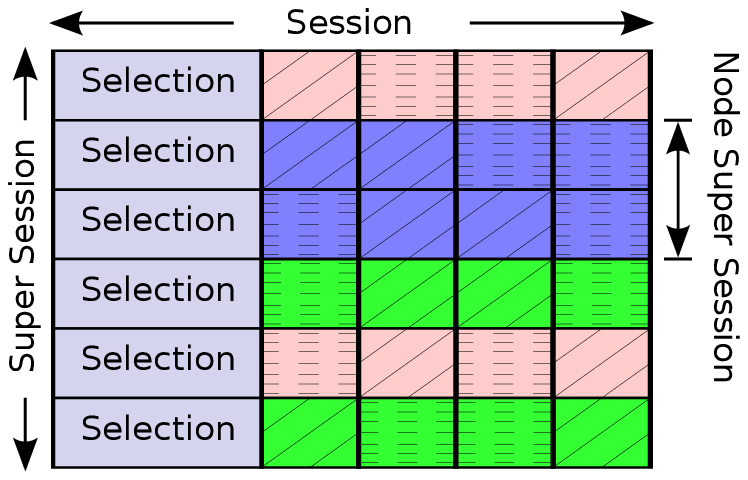}
\label{fig:rn_fd}
}\subfigure[Fair node selection and random direction  division
(FN-RD).]{
\includegraphics[height=0.16\textwidth,width=0.25\linewidth]{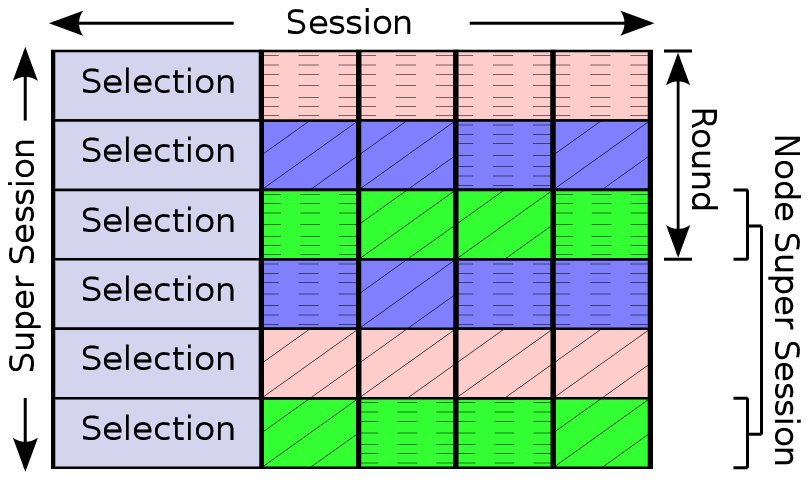}
\label{fig:FN_rd}
}\subfigure[Random node selection and random direction division (RN-RD).]{
\includegraphics[height=0.16\textwidth,width=0.23\linewidth]{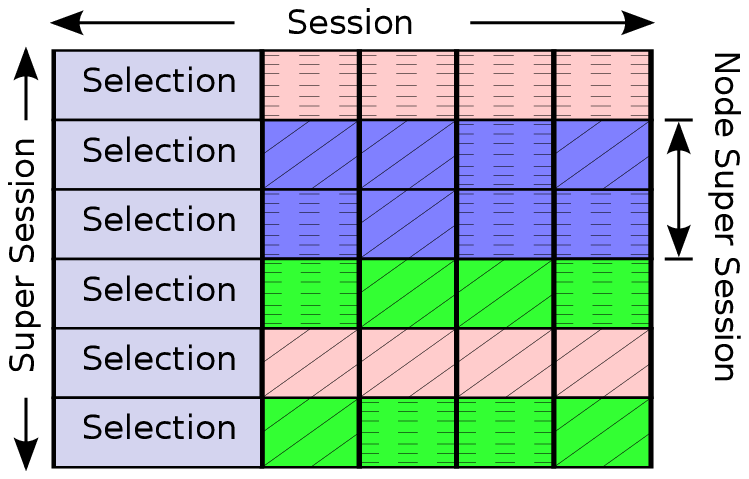}
\label{fig:rn_rd}
}
\caption{Four different schemes for two-way communication. The direction selection phase between the selection and data transmission phase is not shown for clarity. Different colors represent different nodes while different hashing pattern represent the from/to slots.}
\label{fig:extensions}
\end{figure*}

The following notations are illustrated in Figure~\ref{fig:extensions}.
\begin{itemize}
  \item A \textbf{session} is a group of slots that represent a single selection
phase followed by a data transmission phase. The data phase of each session
contains $f$ frames.

  \item A \textbf{round} is defined as a group of $n$ sessions, in which each one of the $n$ nodes is
assigned one session.

  \item A \textbf{supersession} is a group of sessions ($2nk$) in which \emph{all nodes}
finish the transmission of one message (i.e. a transmission of $k$ frames in each direction from/to the coordinator).

  \item A \textbf{node supersession} is a group of $k$ sessions that belong to \emph{one}
node in which this node finishes the transmission of one message.

\end{itemize}

\subsection{Fair node selection and fair direction division scheme (FN-FD)}
This scheme combines short term node fairness and short term
direction fairness (Figure~\ref{fig:FN_fd}). In particular, all nodes must take a turn within the round (in a random fashion) before a node can be assigned another turn by the coordinator. The number of from/to slots within each data transmission session are equal. Therefore, both node and direction distributions have short
term fairness. This scheme, however, reduces the
ambiguity at the eavesdropper and hence decreases security as we quantify in the next section.

\subsection{Random node selection and fair direction division scheme (RN-FD)}

In this scheme (Figure~\ref{fig:rn_fd}), the number of from/to frames within each data session has to be equal (achieving direction fairness). However, the sessions assigned to a specific node can be anywhere within the supersession (random node selection), i.e. there are no rounds.

\subsection{Fair node selection and random direction division scheme (FN-RD)}

In this scheme (Figure~\ref{fig:FN_rd}), each node has to be selected at least once before another node
gets a second chance. In other words, each node will take a turn within the
round. The direction of traffic from/to the coordinator needn't be balanced
within a session, but is balanced on the long term in the supersession.

The send/receive queue at the coordinator though may not be balanced due to the random
direction assignment. The constraints of the long term fairness over the
direction of traffic increases the amount of state that needs to be kept at the
coordinator.

\subsection{Random node selection and random direction division scheme (RN-RD)}
In this last scheme, the coordinator divides the sessions among the nodes and the from/to traffic randomly
within the supersession. Therefore, both node and direction distributions do
not have short term fairness (Figure~\ref{fig:rn_rd}).

This scheme has the advantage of increasing the ambiguity at the
eavesdropper and hence increasing security. However, it lacks short term
fairness and the coordinator has to keep track of more state for the long term
fairness.

\section{Analysis}
\label{sec:analysis}
In this section, we analyze the different schemes through analysis and
simulation in terms of security, overhead, and fairness.
\subsection{Secrecy}

For security, we have two modes, depending on the eavesdropper goal. In the
first mode, (single node) the eavesdropper is only interested in the messages of
a specific node. The second mode assumes that the eavesdropper is interested in
the entire network traffic.

\subsubsection{Single node attack}
Let the selected node for the attack be $s$. We analyze the security of the four
different schemes. Note that the total number of slots for all nodes to transmit
one message in each direction each is $2nk$. The corresponding total number of sessions therefore
is $\frac{2nk}{f}$ and the number of sessions allocated to a single node is
$\frac{2k}{f}$.

\paragraph{\textbf{RN-FD scheme}}
In order for Eve to guess the message of $s$, it needs to guess the sessions
assigned to $s$. This occurs with probability

\begin{equation}
=\frac{1}{\binom{\frac{2nk}{f}}{\frac{2k}{f}}}
\label{R_S}
\end{equation}

In addition, Eve has to guess the direction of the frames to avoid mixing the
packets from/to the coordinator. This occurs with probability:
\begin{equation}
={\frac{1}{\binom{f}{\frac{f}{2}}^{\frac{2k}{f}}}}
\label{F_T}
\end{equation}

Therefore, the outage probability for this scheme, for an $n$-node network
($\Pr_{\textrm{out}}(n)$) is:
\begin{equation}
Pr_{\textrm{out}}(n)=\left.\frac{1}{\binom{\frac{2nk}{f}}{\frac{2k}{f}}}\right.
{\frac{1}{\binom{f}{\frac{f}{2}}^{\frac{2k}{f}}}}
\end{equation}

\paragraph{\textbf{FN-RD scheme}}
In this case, Eve needs to decide in each round which session belongs to $s$.
Therefore the probability of correctly guessing Eve's sessions in the entire
supersession is:

\begin{equation}
={\left(\frac{1}{n}\right)^{\frac{2k}{f}}}
\label{3R_S}
\end{equation}

Once the sessions of node $s$ are determined, Eve has to guess which of the total of $2k$
frames are to the coordinator and which are from it. This occurs with
probability:

\begin{equation}
={\frac{1}{\binom{2k}{k}}}
\label{R_T}
\end{equation}

Therefore, the outage probability is this case is
\begin{equation}
Pr_{\textrm{out}}(n)=\left.{\left(\frac{1}{n}\right)^{\frac{2k}{f}}}\right.{
\frac{1}{\binom{2k}{k}}}
\end{equation}

\paragraph{\textbf{RN-RD scheme}}

Similar to equations \ref{R_S} and \ref{R_T}, the outage probability in this
case is:

\begin{equation}
Pr_{\textrm{out}}(n)=\left.\frac{1}{\binom{\frac{2nk}{f}}{\frac{2k}{f}}}\right.
{\frac{1}{\binom{2k}{k}}}
\end{equation}

\paragraph{\textbf{FN-FD scheme}}

Similarly, the outage probability here can be obtained by combining equations
\ref{F_T} and \ref{3R_S} as:

\begin{equation}
Pr_{\textrm{out}}(n)=\left.{\left(\frac{1}{n}\right)^{\frac{2k}{f}}}\right.{
\frac{1}{\binom{f}{\frac{f}{2}}^{\frac{2k}{f}}}}
\end{equation}

\subsubsection{Network-wide attack}
In this attack, Eve is interested in obtaining the entire network traffic. Once
Eve guesses the frames of one node, the problem size decreases to that of an $n-1$
node network. Therefore, the outage probability in this case
($Pr_{\textrm{NOut}}$) is:

\begin{equation}
Pr_{\textrm{NOut}} =\prod_{i=n}^{1} Pr_{\textrm{out}}(i)
\end{equation}

\subsection{Overhead}
The four different schemes have the same overhead which is due to the selection
and direction determination phases. Therefore, the overhead for all four schemes is:

\begin{equation}
\frac{t(n+f)}{fl+t(n+f)}
\end{equation}

which is a function of the number of nodes ($n$), dialog codes preamble length ($t$), number of slots in a session
($f$), and the data frame length ($l$).

\begin{figure*}[!t]
\centering
\subfigure[Num. of nodes]{
\includegraphics[width=0.31\linewidth]{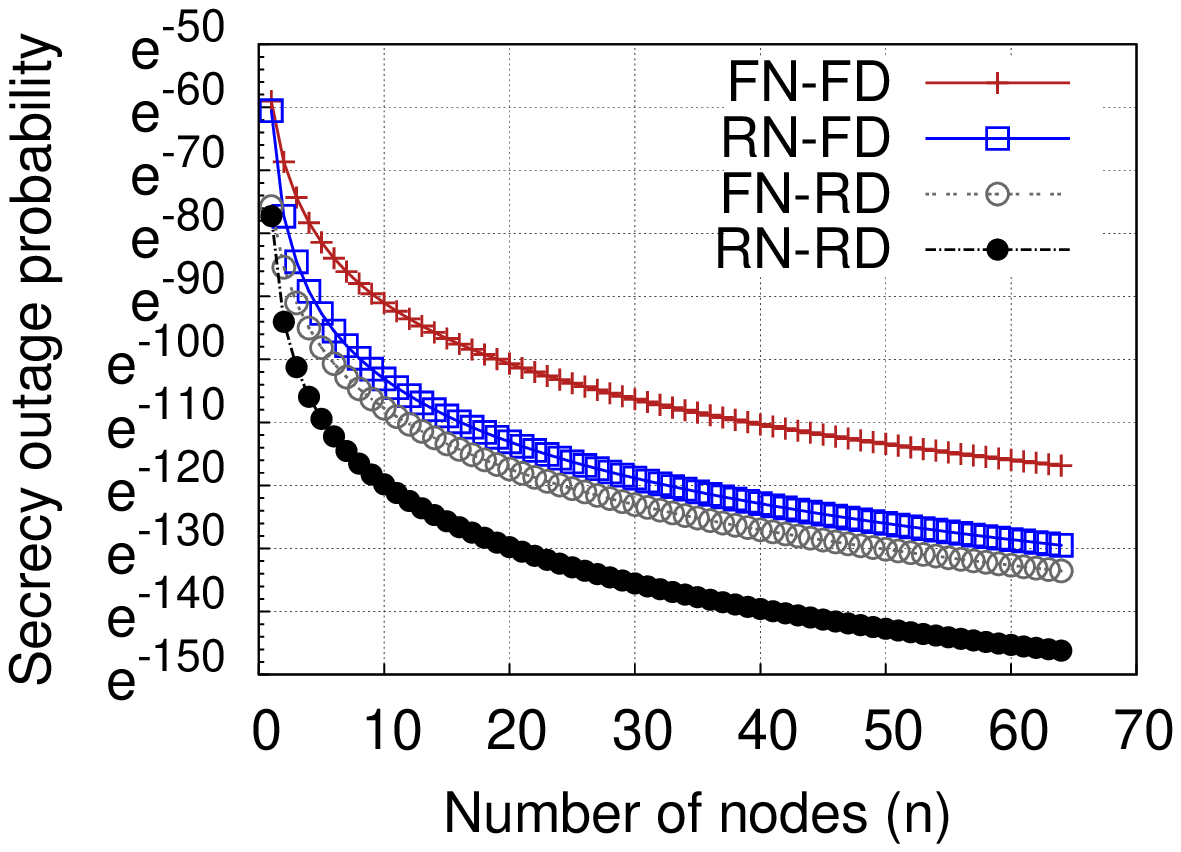}
}
\subfigure[Num. of frames/message]{
\includegraphics[width=0.31\linewidth]{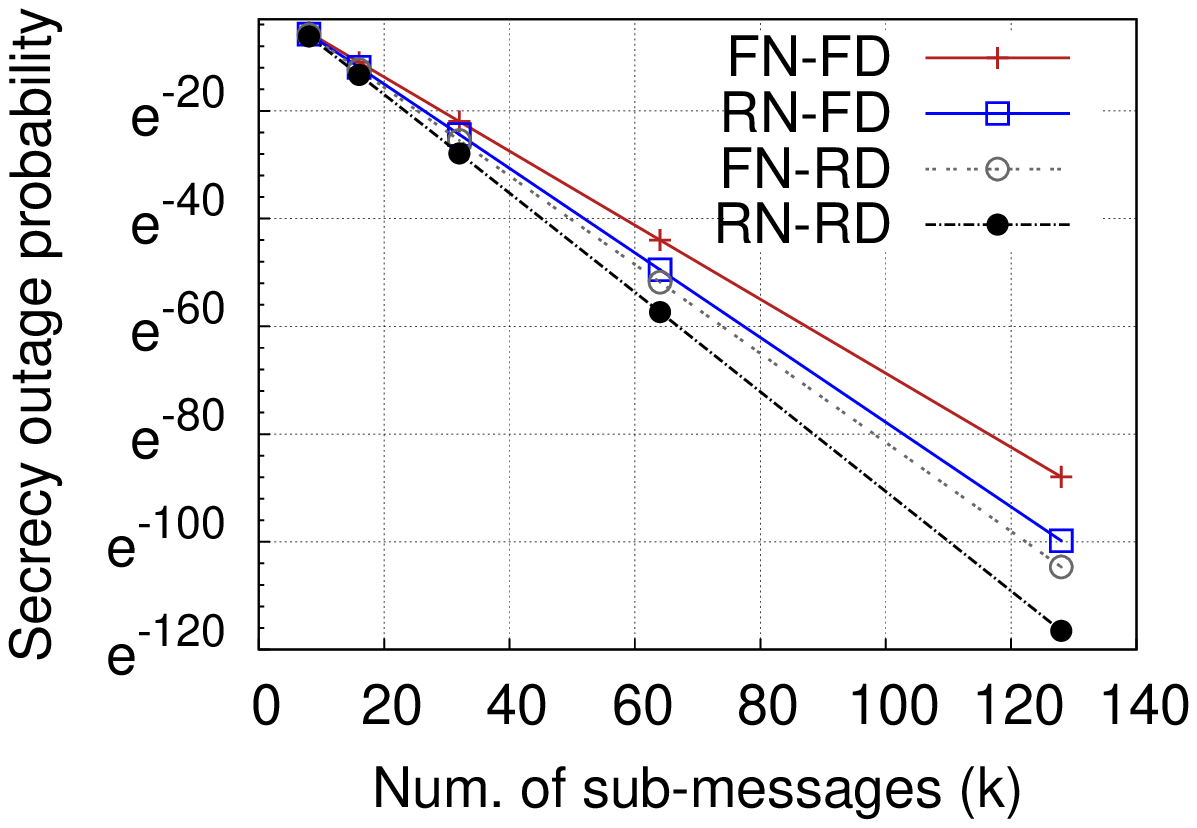}
}
\subfigure[Session length]{
\includegraphics[width=0.31\linewidth]{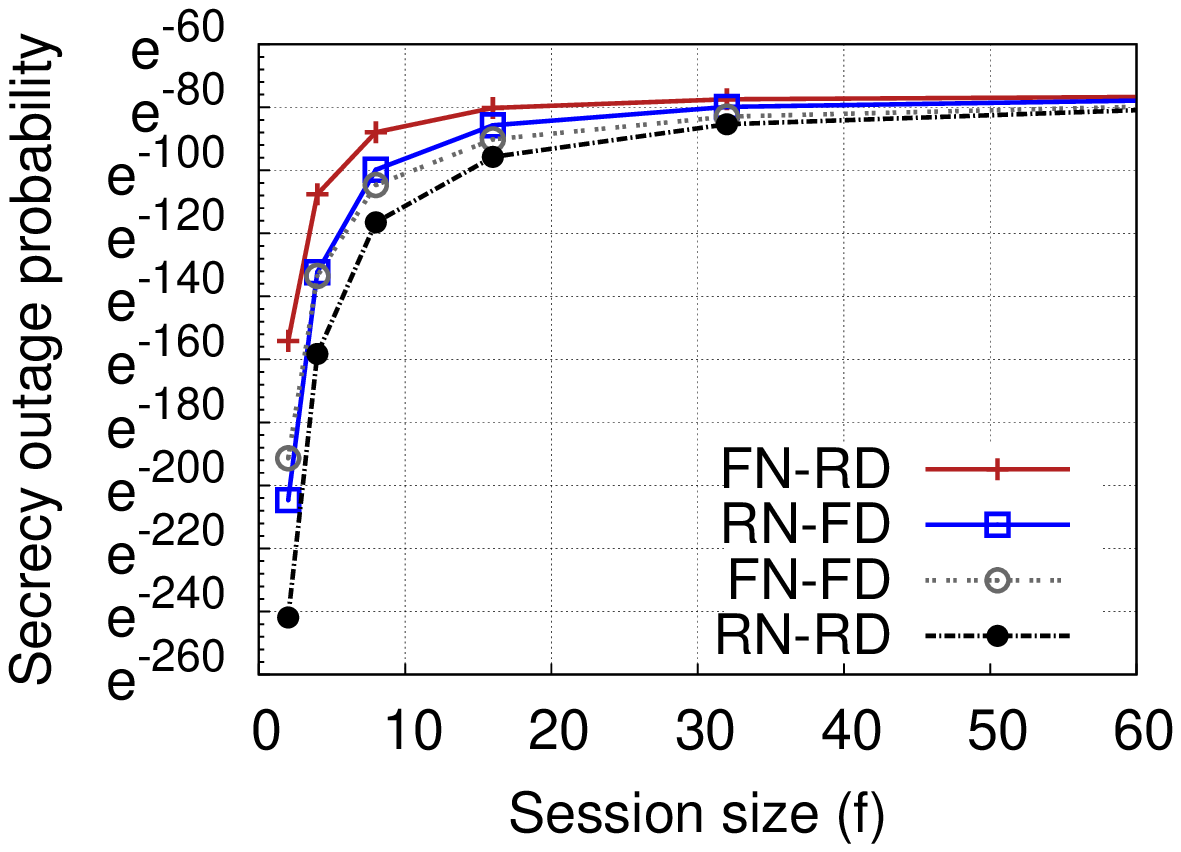}
}
\caption{Effect of different parameters on secrecy.}
\label{fig:secrecy}
\end{figure*}

\subsection{Numerical Results}
\subsubsection{Secrecy analysis}
Figure~\ref{fig:secrecy} shows the effect of changing the system parameters,
i.e. $n$, $k$, and $f$, on the outage probability for the four schemes. The
figure shows that all schemes have the advantage of enhancing the secrecy with
the increase of the number of nodes in the network. Increasing $k$ increases the
space of guessing at Eve, and hence enhances secrecy. Increasing $f$ leads
to increasing the length of the data transmission phase and hence reducing the
frequency of the selection phase. This reduces secrecy.

The figure also shows that the RN-RD scheme has the highest secrecy. This is due
to the increased ambiguity at Eve due to the randomization of both node
selection and direction. On the other extreme, the FN-FD scheme has the least secrecy.
The other two schemes have a secrecy outage probability in between: As the data
phase length increases, direction randomization leads to more secrecy than node
randomization.

\subsubsection{Overhead}
\begin{figure}[!t]
\centering
\includegraphics[width=0.7\linewidth]{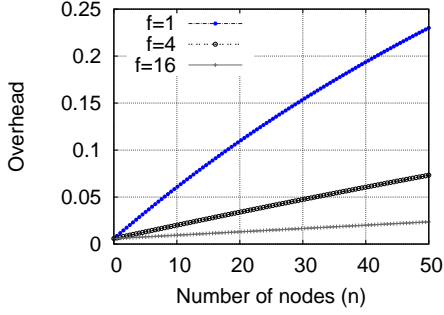}
\caption{Effect of different parameters on overhead ($l= 1024$ bits, $t=6$). Note that a longer frame length ($l$) leads to lower overhead.}
\label{fig:overhead}
\end{figure}

Figure~\ref{fig:overhead} shows the effect of changing the system parameters on
the system overhead. The figure shows that the overhead increases with
the increase of the number of nodes in the network and the decrease of the data
transmission phase length. Therefore, a trade-off exists between overhead and
secrecy. The operation point can be selected based on the specific application
need.

\subsubsection{Fairness}
\begin{figure}[!t]
\centering
\subfigure[Num. of nodes]{
\includegraphics[width=0.46\linewidth]{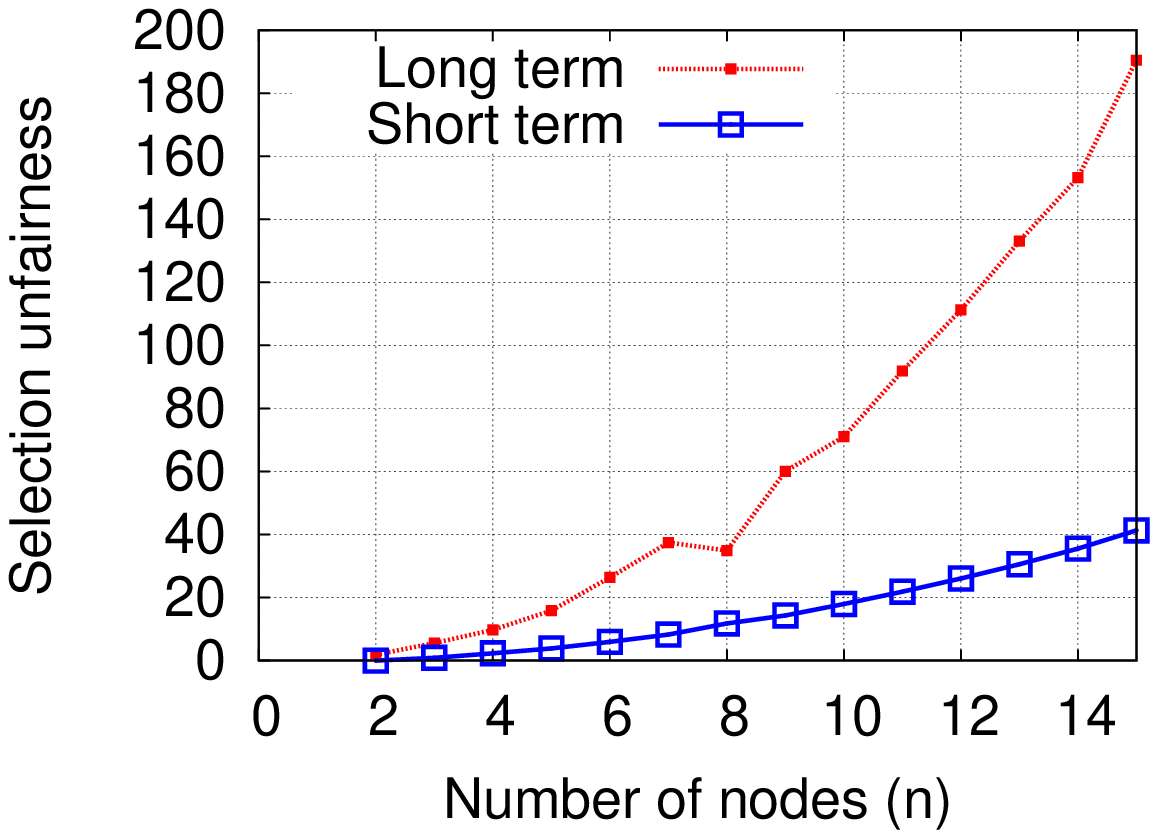}
}
\subfigure[Num. of sessions]{
\includegraphics[width=0.46\linewidth]{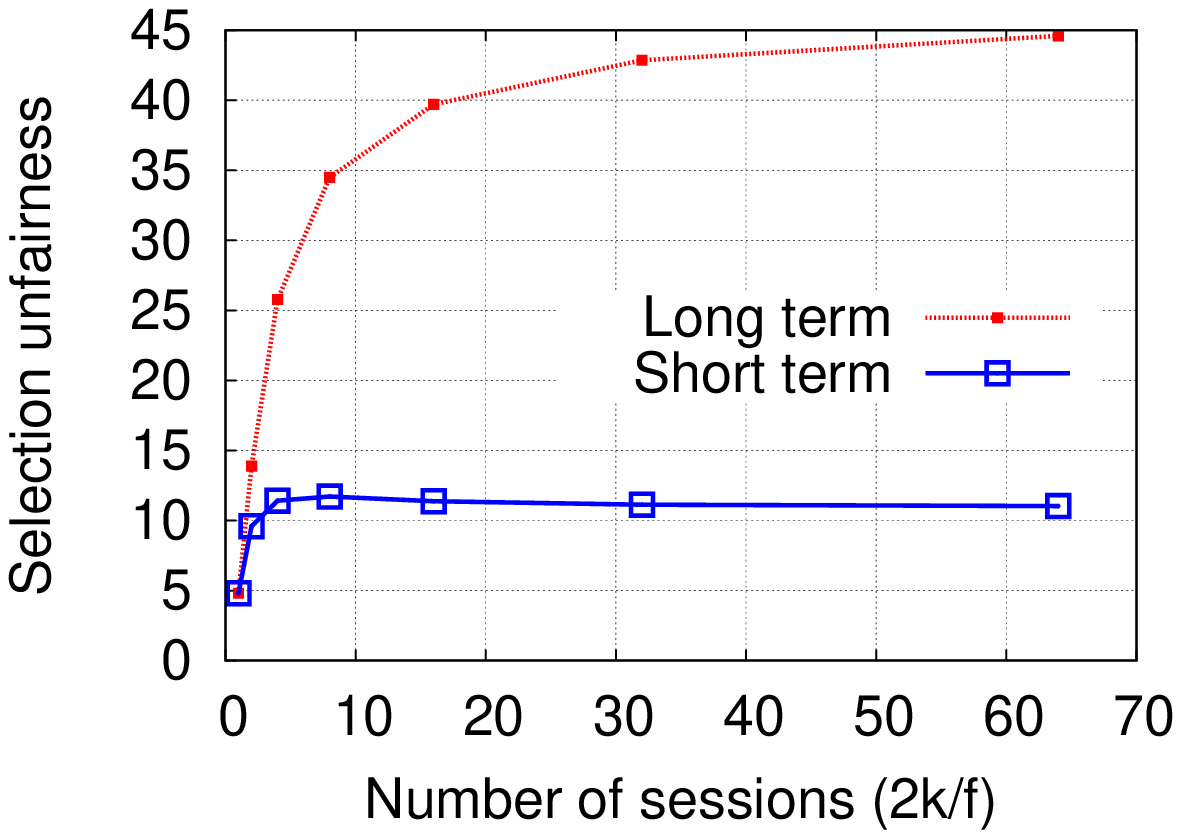}
}
\caption{Effect of different parameters on node fairness.}
\label{fig:fairness_node}
\end{figure}

For the fairness in node selection, we use the variance of the \emph{difference
between two consecutive sessions} indices as our metric. The more consistent
this difference, the lower the variance, and the higher the fairness. More
formally, if the session indices assigned to a node are $\{s_i\| 0<i< 2k/f\}$,
then $d_i= s_{i+1}- s_i$ and the unfairness index equals $\textrm{Var}(d_{i})$.
Figure~\ref{fig:fairness_node} shows the effect of the different parameters on
node fairness. The figure confirms that the round-based schemes are fairer than
the random schemes. As the number of nodes ($n$) increases, the unfairness
increases. On the other hand, for a fixed $n=8$, increasing the number of
sessions, by either increasing $k$ or reducing $f$, the unfairness increases.
However, this is limited to within a round in the short term node fairness (FN) schemes
and is more variable in the long term node fairness schemes. The saturation in both
cases is due to the limitation imposed by the supersession size.

For direction fairness, our metric is the absolute difference between the sum of
the send and receive indices within a specific node supersession, averaged over
all nodes. The smaller this number, the higher the fairness. Note that since the
fairness metric is node based, it is independent from the number of nodes $n$.

\begin{figure}[!t]
\centering
\subfigure[Num. of sub-messages]{
\includegraphics[width=0.46\linewidth]{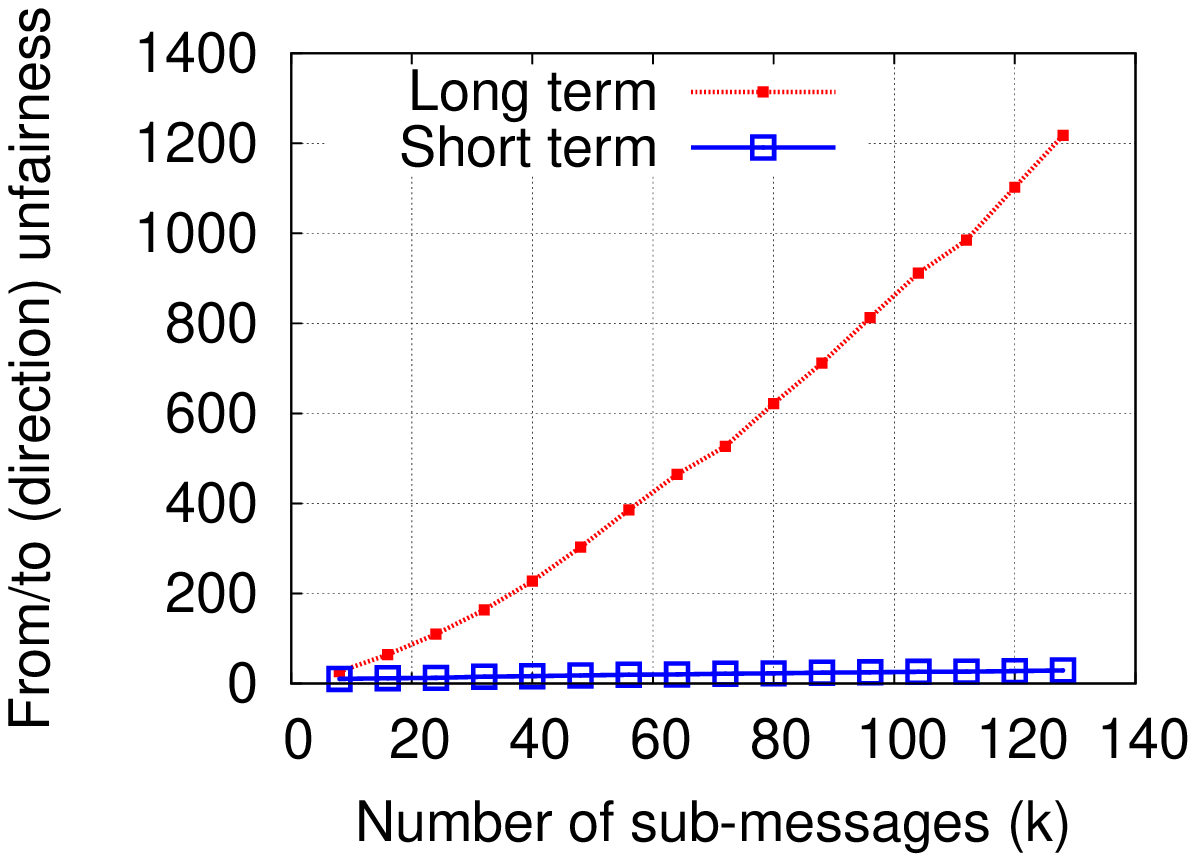}
}
\subfigure[Session size]{
\includegraphics[width=0.46\linewidth]{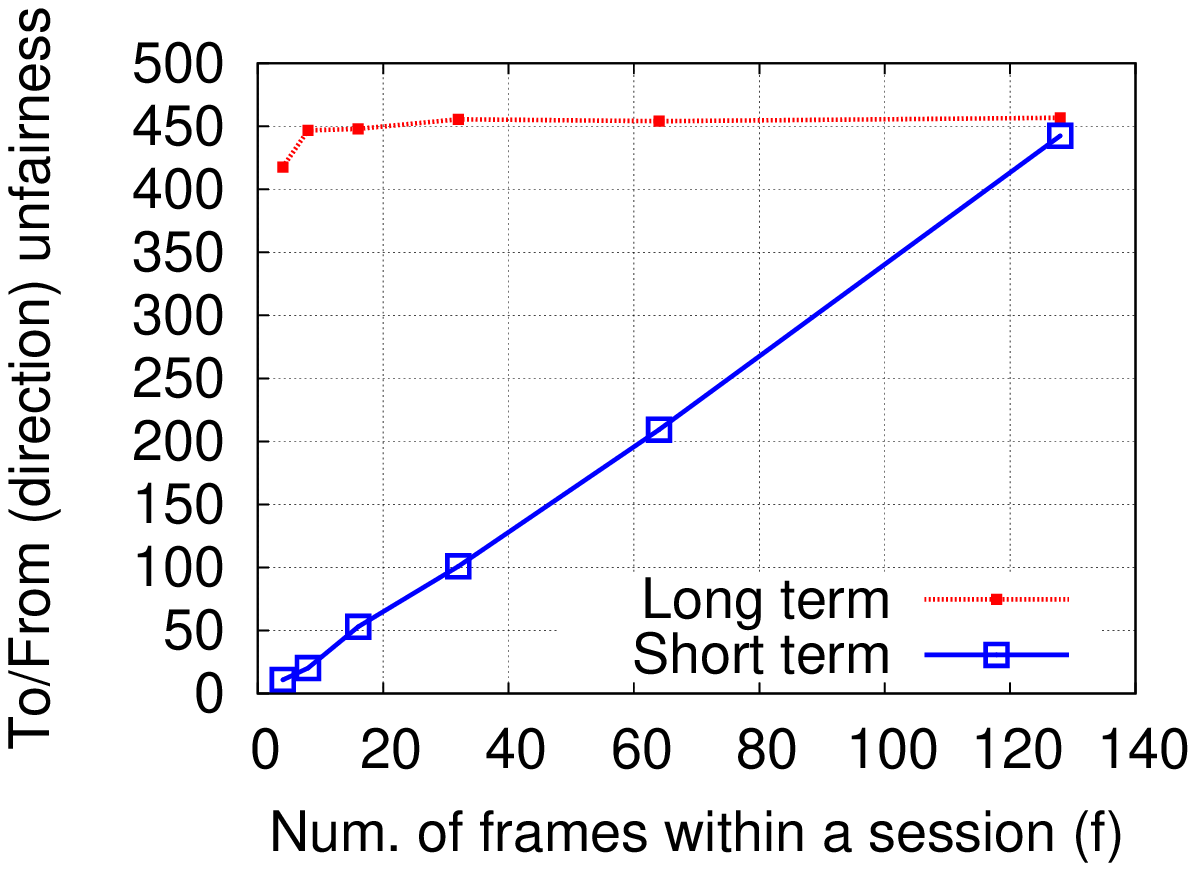}
}
\caption{Effect of different parameters on direction fairness.}
\label{fig:fairness_dir}
\end{figure}

Figure~\ref{fig:fairness_dir} shows the effect of the different parameters on
direction fairness. The figure confirms that the short term direction fairness (FD) schemes are fairer
than the long term direction fairness (RD) schemes. As the number of frames required to construct a mesage  ($k$) increases, the
unfairness increases in the long term direction fairness scheme as the overall number of
slots in the node supersession will increase. $k$ has no effect on the short term direction fairness schemes as all direction selections are based on a round, which
is independent of $k$. This is the opposite case as we fix $k$ and change the
number of frames within a session ($f$). In this case, the performance of the
completely random case is independent of the number of frames within a session,
as all sessions are concatenated in one supersession. Increasing $f$ increases
the unfairness of the short term direction fairness schemes. However, their worst case
performance is bounded by the performance of the long term direction fairness scheme, where
on session becomes a node supersession.

\subsubsection{Comparison with a traditional two-node system}
\begin{figure}[!t]
  \centering
  \subfigure[Secrecy]{
  \includegraphics[width=0.46\linewidth]{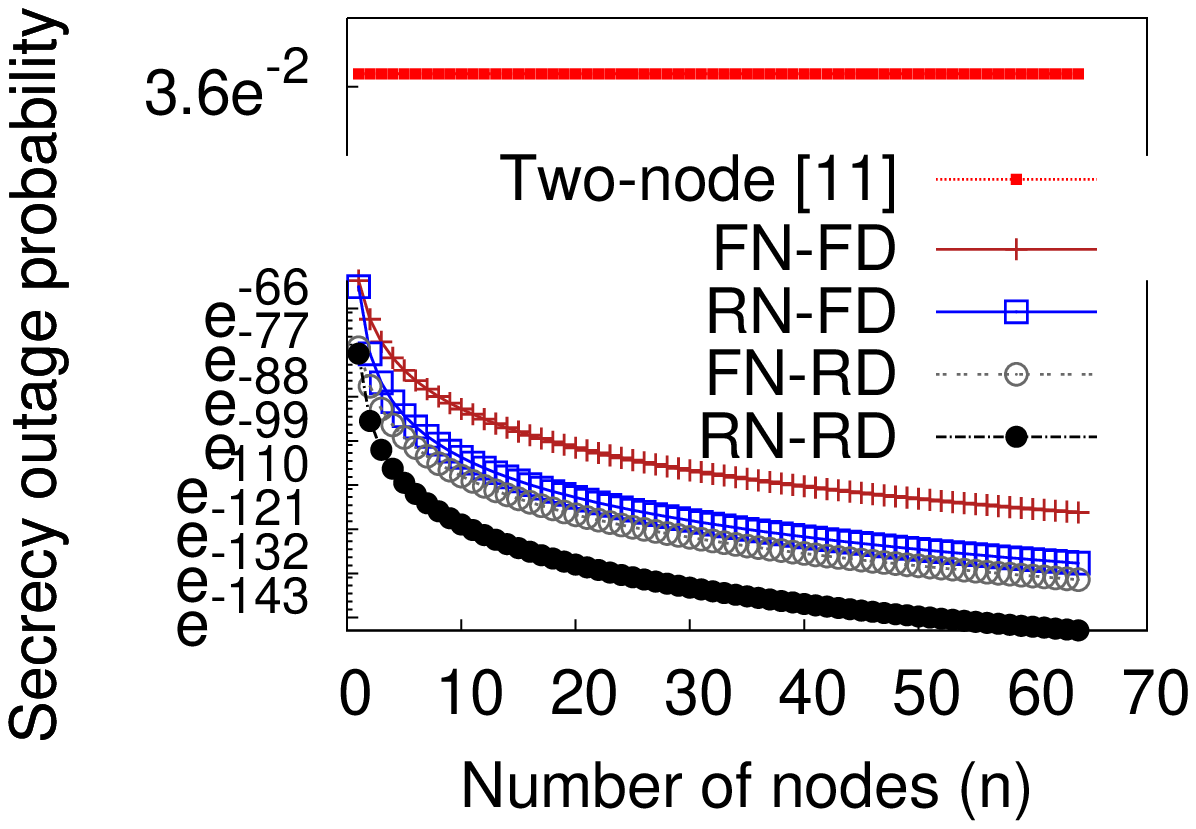}
  }
  \subfigure[Overhead]{
  \includegraphics[width=0.46\linewidth]{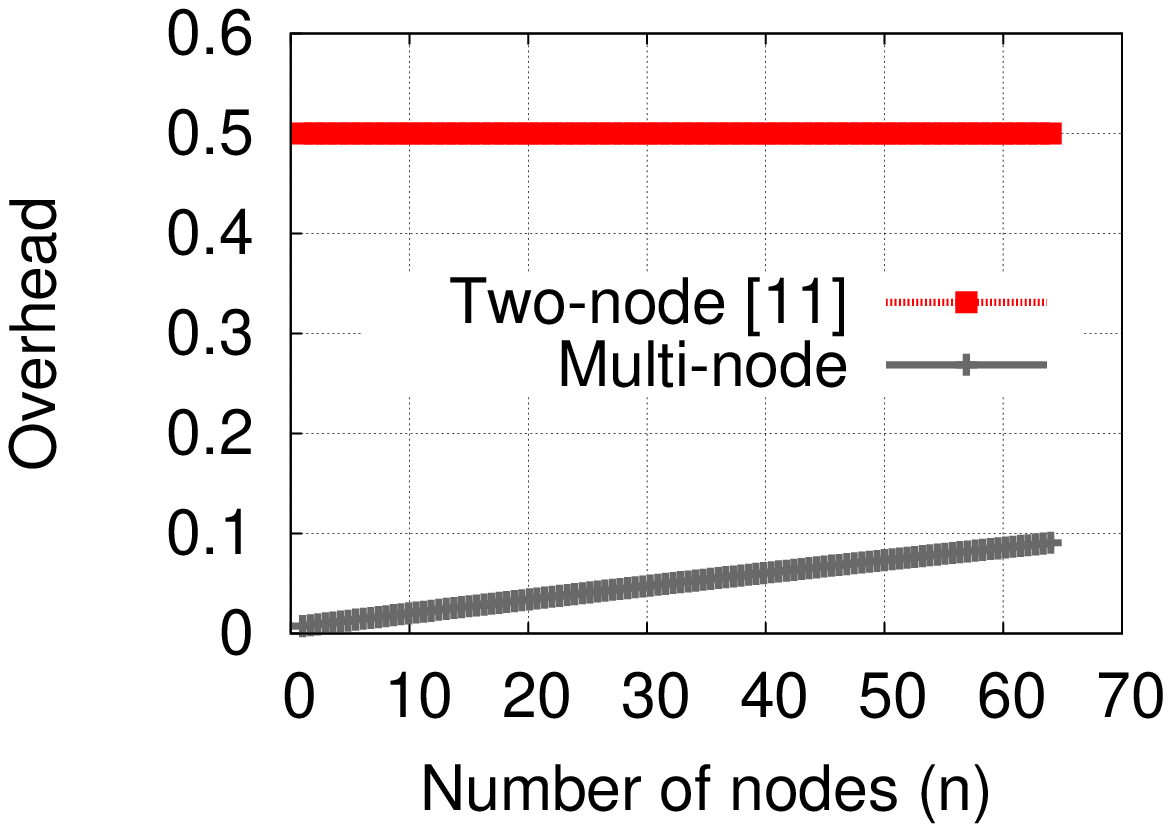}
  }
\caption{Comparison with previous two-node based practical provably secure
schemes \cite{Elmorsy}.}
  \label{fig:compare}
\end{figure}

Figure~\ref{fig:compare} compares the proposed schemes to the practical provably
secure two-node scheme proposed in \cite{Elmorsy} under typical parameters for
all schemes. The scheme in \cite{Elmorsy} is based on randomization between two
nodes. A direct extension for this case to the multi-node case is to apply it
pairwise to each transmitter receiver. The figure shows that this reduces
secrecy significantly, with several orders of magnitude and this loss in secrecy
increases with the increase in the number of nodes. Since \cite{Elmorsy} does
not leverage the multi-node in its design, both its security and overhead is
independent of $n$. Our proposed schemes also have much better overhead under
typical network sizes.

\subsection{System Implementation}
We have also implemented the proposed scheme on TelosB motes equipped with CC2420 radio chips which come with
half-duplex antennas. The motes run the TinyOS operating system.

The network consists of three types of nodes nodes:

\begin{enumerate}
\item An observer node, which plays a double role in our setup. First, it
plays the role of the passive eavesdropper which sniffs all sent frames.
Second, it is responsible for synchronizing all nodes in the network by sending
a pulse (synchronization frame in our case) at constant intervals to initiate the
start of a slot and hence transmission of frames. The remaining nodes react to
 these synchronization frames.

\item A normal node (representing one of the $n$ legitimate nodes).

\item A coordinator node, which selects which node to transmit and the direction of traffic.
\end{enumerate}

The implementation results confirm the analysis results in the previous sections. More details about the implementation can be found in \cite{multinode_demo}.

\section{Conclusion}
\label{sec:conclude}
We presented a novel practical and provably secure solution to the multi-node wireless communication problem. Our solution is based on hiding the identity of the communicating nodes from the eavesdropper. We presented four different variations of the basic scheme that can achieve different fairness-security tradeoffs.
We evaluated the proposed techniques using analysis and implementation. Our results show that our scheme outperforms direct extensions of the two-node communication schemes in terms of both overhead and secrecy, highlighting its suitability for highly secure applications.

\bibliographystyle{IEEEtran}

\end{document}